\documentstyle[aps]{revtex}
\begin{document}
\title{ANKARA\ UNIVERSITY FACULTY\ OF\ SCIENCES\hspace{1.0in}\hspace{1.0in}
DEPARTMENTS OF PHYSICS AND ENGINEERING\ PHYSICS\\
\smallskip\ \\
\hspace{1.0in}\hspace{1.0in}\hspace{1.0in} \hspace{1.0in}\hspace{1.0in}%
AU-HEP-99/01\\
\hspace{1.0in}\hspace{1.0in}\hspace{1.0in}\hspace{1.0in}\hspace{1.0in}9 June
1999\\
\vspace{1.0in}Linac-Ring Type $\phi $ Factory for Basic and Applied
Researches}
\author{A.K. \c {C}ift\c {c}i$^a$, O.G\"{u}rkan$^b$, T. Olgar$^b$, E. Recepoglu$^a$,
S. Sultansoy$^{a,c}$, \"{O}. Yavas$^b$, M. Yilmaz$^d$}
\address{$^{a)}$ Department of Physics, Faculty of Sciences, Ankara University, 06100%
\\
Tandogan-Ankara, TURKEY\\
$^{b)}$ Department of Engineering Physics, Faculty of Sciences, Ankara\\
University, 06100 Tandogan-Ankara, TURKEY\\
$^{c)}$ Institute of Physics, Academy of Sciences, Cavid av. 33, Baku,\\
AZERBAIJAN\\
$^{d)}$ Department of Physics, Faculty of Arts and Sciences, Gazi\\
University, 06500 Besevler-Ankara, TURKEY}
\maketitle

\begin{abstract}
In this paper, main parameters for linac-ring type collider desinged for
producing $\phi $ particles copiously are estimated and potential of this
machine in particle physics researches is investigated. Moreover, parameters
for free electron laser and synchrotron radiation obtained from electron
linac and positron ring, respectively, are determined and applications of
these radiations are summarized.
\end{abstract}

\section{INTRODUCTION}

The center of mass energy $\sqrt{s}$ needed to produce $\phi $ particles is
enough to be about 1 GeV. In this study main parameters of two linac-ring
type collider options are given: one with 125 MeV linac electron beam and 2
GeV positron beam, another with 250 MeV linac electron beam and 1 GeV
positron beam. Main reason for designing collider as linac-ring type machine
is the possibility to increase luminosity by of one or two orders with
respect to standart $\phi $ factories. Today the highest luminosity among
the standard (ring-ring type) $\phi $ factories is owned by $DA\phi NE$ with 
$L=5\times 10^{32}cm^{-2}s^{-1}$\cite{dafne}. Proposed collider will give
opportunity to reach luminosity $\sim 10^{34}cm^{-2}s^{-1}$.

Great number of neutral and charged $K$ mesons ($>10^{11}$ per working
year), produced as a result of decays of $\phi $ mesons, can be
investigated. The importance of K mesons study is obvious: for example, the
CP-violation (matter-antimatter asymmetry) was observed and established in
neutral K mesons' decays.

Proposed accelerator complex will give opportunity to perform a large
spectrum of applied researches: positron ring can be used as the third
generation synchrotron radiation (SR) source and Free Electron Laser (FEL)
can be constructed on the base of main electron linac.

In following section we present general overwiev of linac-ring type $\phi $
factory. Main parameters of proposed machine are estimated in Section III.
Then, in Section IV physics search potential of the collider is briefly
discussed. Parameters of SR and FEL photon beams are estimated in Sections V
and VI, respectively. We also list the possible applications of these beams
in different fields of science and technology in Section VII. Finally, in
Section VIII we give some concluding remarks.

\section{GENERAL\ OVERWIEV}

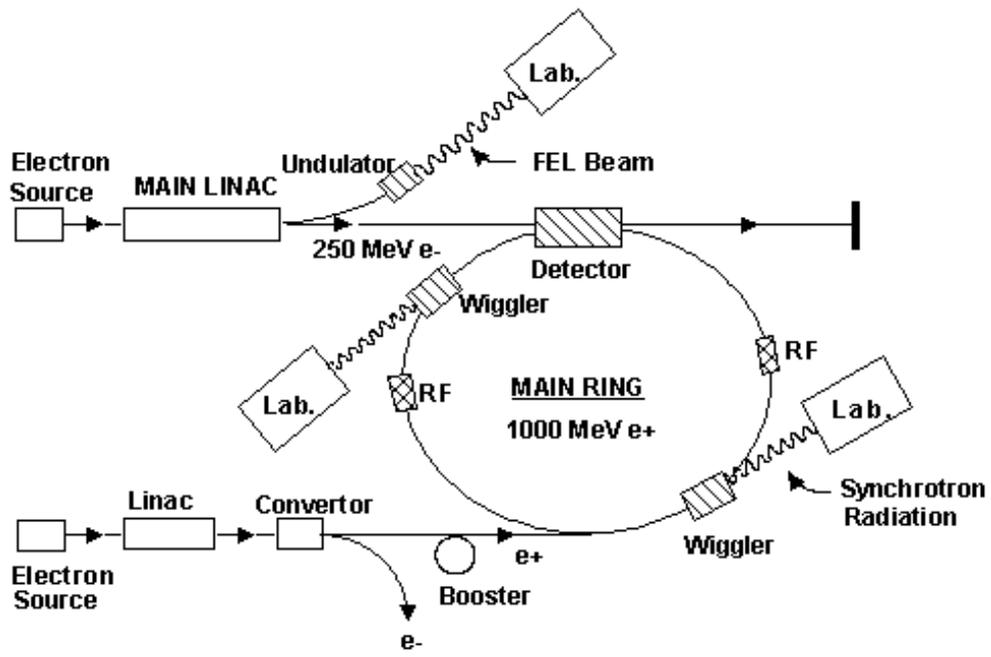
\begin{figure}[tbp] \centering
\caption{The general scheme of
the proposed complex.\label{fig.s.1}}%
\end{figure}

The general scheme of proposed complex is given in Fig. \ref{fig.s.1}.
Electrons accelerated in main linac up to energies 250 (125) MeV are
forwarded to detector region where they collide with positrons from main
ring or turned out to undulator region where FEL beam is produced.

On the other side electrons, accelerated in small linac, are forwarded to
conversion region where positron beam is produced. Then, positrons are
accumulated in booster and after some beam gymnastics are forwarded to the
main ring and accelerated up to energies 1(2) GeV. Wigglers installed in two
regions will provide SR for applied researches.

\section{MAIN\ PARAMETERS\ OF\ LINAC-RING\ TYPE $\phi $ FACTORY}

The usage of linac-ring type colliders as a particle factories is widely
discussed during the last decade. Below we list some proposals with
corresponding references:

i) B Factory \cite{bfac},

ii) c-$\tau $ Factory \cite{cfac},

iii) Z Factory \cite{zfac}.

The main advantages of linac-ring type machines are: the possibility to
achieve essentially higher luminosities with respect to standard ring-ring
type particle factories and asymmetric kinematics. Of course, LR type $\phi $
factory is most compact one because of lowest center-of-mass energy.

Main parameters of proposed machine are given in Table \ref{table.s.1} for
two different choices of electron and positron beam energies. Below we
present several illuminating notes.

Electron bunches accelerated in main linac are used only once for
collisions. On the other hand, positron bunches have to be used numerously,
therefore, the stability of positron beam is very important. The condition
of stability is given by

\begin{equation}
n_{e^{-}}=8.7\times 10^8\cdot E_{e^{+}}\left[ GeV\right] \cdot (\sigma
\left[ \mu m\right] )^2\cdot (\beta ^{*}\left[ cm\right] )^{-1}\cdot \Delta Q
\end{equation}
where $n_{e^{-}}$ is number of particles in electron bunch, $E_{e^{+}}$ is
positron beam energy, $\sigma $ is transverse size of beams and $\beta ^{*}$
is amplitude function at the collision point, $\Delta Q$ is tune shift
caused by collision. Emphrically, $\Delta Q\leq 0.06$ for lepton beams
storaged in rings. In principle, this upper limit taken from experiments
done in usual ring-ring type $e^{+}e^{-}$ colliders can be higher for
linac-ring type machines. In this paper we use the conservative value $%
\Delta Q\leq 0.06$.

Synchrotron radiation resulted from bending magnets used in ring type
accelerators causes to decrease in beam energy. Energy loss happened as a
result of this radiation in every tour is given by

\begin{equation}
\Delta E_{e^{+}}[MeV]=0.0885\cdot (E_{e^{+}}\left[ GeV\right] )^4\cdot
(R\left[ m\right] )^{-1}\text{ }
\end{equation}
where $R$ is radius of the ring.

The fractional energy loss of electrons in positron beam field is given by

\begin{equation}
\delta =(n_{e^{+}}\left[ 10^{12}\right] )^2\cdot (\sigma _z\left[ cm\right]
)^{-1}\cdot (\sigma _x\left[ \mu m\right] \cdot \sigma _y\left[ \mu m\right]
)^{-1}\cdot E_{e^{+}}\left[ TeV\right]
\end{equation}
where $n_{e^{+}}$ is number of particles in one positron bunch, $\sigma
_{x,y}$ are vertical and horizontal transverse sizes of beam at the
collision point (in our case $\sigma _x=\sigma _y=\sigma $), $\sigma _z$ is
the bunch length.

Electron current is given by

\begin{equation}
I_{e^{-}}\left[ mA\right] =1.6\times 10^{-19}\cdot n_{e^{-}}\cdot f
\end{equation}
where $f$ is the collision frequency. Positron current in the ring is

\begin{equation}
I_{e^{+}}\left[ A\right] =1.6\times 10^{-19}\cdot k\cdot n_{e^{+}}\cdot
(c/2\pi R)
\end{equation}
where k is the number of positron bunches in the ring and c is the speed of
light.

\section{PHYSICS SEARCH POTENTIAL}

Quantum numbers of $\phi $ mesons produced as a resonance in $e^{+}e^{-}$
collisions are $I^G(J^{PC})=0^{-}(1^{--}).$ Mass is $m_\phi =1015.413\pm
0.008$ MeV and total decay width is $\Gamma =4.43\pm 0.05$ MeV \cite{pdg}.
Fundamental decay channels and branching ratios are given in Table \ref
{table.s.2}.

Since deviation of the center-of-mass energy of $e^{+}e^{-}$ collisions is
smaller than the total decay width of $\phi $ meson, cross-section in the $%
\phi $ resonance region can be taken as

\begin{equation}
\sigma =(12\pi /m^2)\cdot (\Gamma _c/\Gamma )\simeq 4.4\times 10^{-30}cm^2
\end{equation}
In the proposed complex $4.4\times 10^{11}\phi $ meson, $2.2\times 10^{11}$ $%
K^{+}K^{-}$ pairs and $1.5\times 10^{11}$ $K_L^0K_S^0$ pairs can be produced
in a working year $(10^7s)$. Fundamental problems of particle physics such
as CP violation, rare decays of K mesons etc. can be investigated with
highest statistics. Moreover, kinematical asymmetry can be adventageous for
measuring neutral K meson's oscilations and CP violation parameters.
Detailed analysis of physics search potential will be done in forthcoming
publications.

\section{SYNCHROTRON RADIATION FACILITY}

Charged particles emit electromagnetic radiation when they accelerated in
the circular orbit. This radiation is called synchrotron radiation.
Synchrotron radiation is usually disturbing phenomenon since it causes to
loss of energy of particles. However, synchrotron radiation has very wide
applications since it covers a wide spectrum including x-ray region. Energy
loss with synchrotron radiation is proportional to $\gamma ^4,$ where $%
\gamma $ is the Lorentz factor. To change the spectrum of the radiation,
either synchrotron ring radius or energy of the positrons moving on the ring
should be changed, but both of them are not practical methods. Therefore,
photons with higher energy can be produced by using a series of alternating
directional equal dipole magnets, called wiggler. By inserting wigglers on
the straight parts of the main ring of $\phi $ factory, one can produce
synchrotron radiation for applied researches.

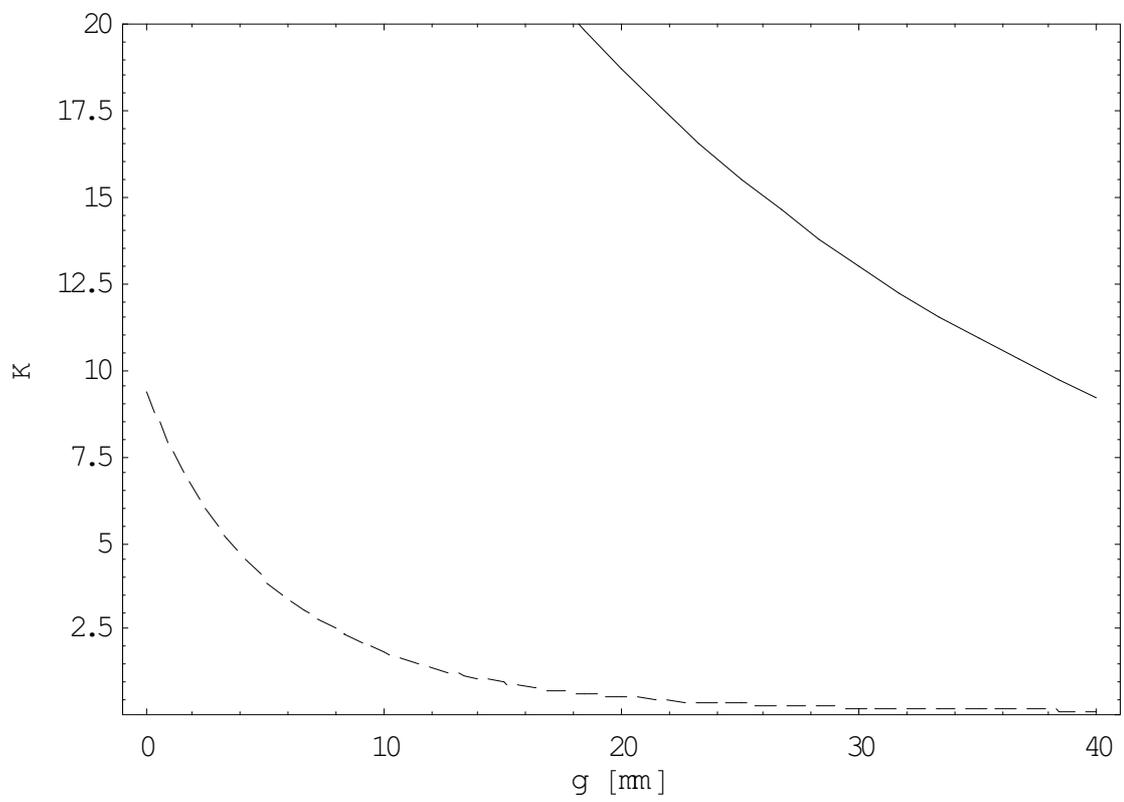
\begin{figure}[tbp] \centering
\caption{The g dependence of
strength parameter for $\lambda_p =33 $ mm  (Dashed line) and   $\lambda_p =132 $ mm  (Solid line). \label{fig.k.1}}%
\end{figure}

When one thinks of whole wiggler, every pole end is designed to effect
particle path neutrally. Photon flux is proportional to number of the magnet
poles. Strength parameter of the wiggler magnet is given by

\begin{equation}
K=0.934\cdot B_0[T]\cdot \lambda _p[cm]
\end{equation}
where $\lambda _p$ is the length between sequential same directional magnet
poles. $B_0$ is the maximum magnetic field strength on midplane axes and its
value for hybrid permanent magnet system is approximately (for $g\leq
\lambda _p$)

\begin{equation}
B_0\simeq B_m\exp [-\frac g{\lambda _p}(b-c\frac g{\lambda _p})]
\end{equation}
where $g$ is vertical distance between magnets, $B_m$ is peak value of
magnet's field, b and c are constants related to used permanent magnets. If
one use SiCo type magnet: $B_m=3.33$ Tesla, $b=5.47$ and $c=1.8$ \cite{wied}%
. Fig. \ref{fig.k.1} shows g dependence of strength parameter of the wiggler
for two values of $\lambda _p.$

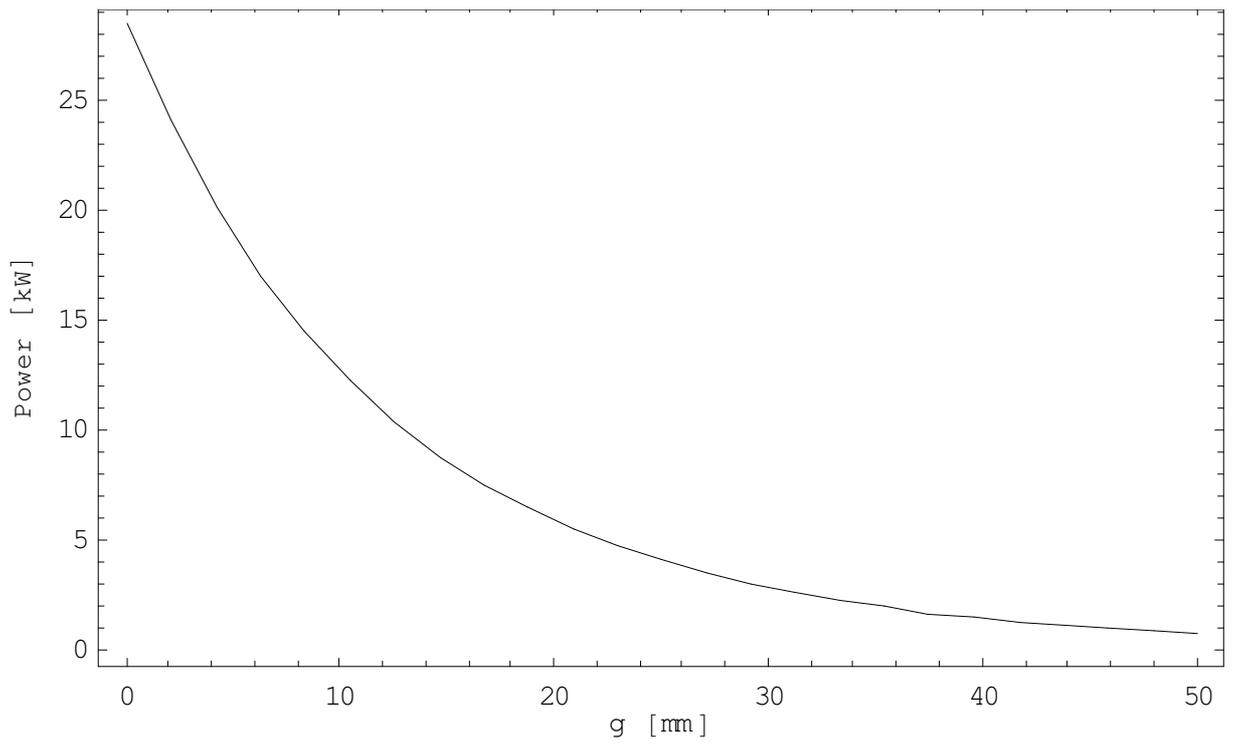
\begin{figure}[tbp] \centering
\caption{The g dependence of
the SR power.\label{fig.k.2}}%
\end{figure}

Power emitted by the wiggler is given by 
\begin{equation}
P[kW]=0.632\cdot L[m]\cdot I_{e^{+}}[A]\cdot (E_{e^{+}}[GeV])^2\cdot
(B_0[T])^2
\end{equation}
where $L$ is the total length of wiggler. The power of the designed
wiggler's radiation with respect to g is shown in Fig. \ref{fig.k.2}.

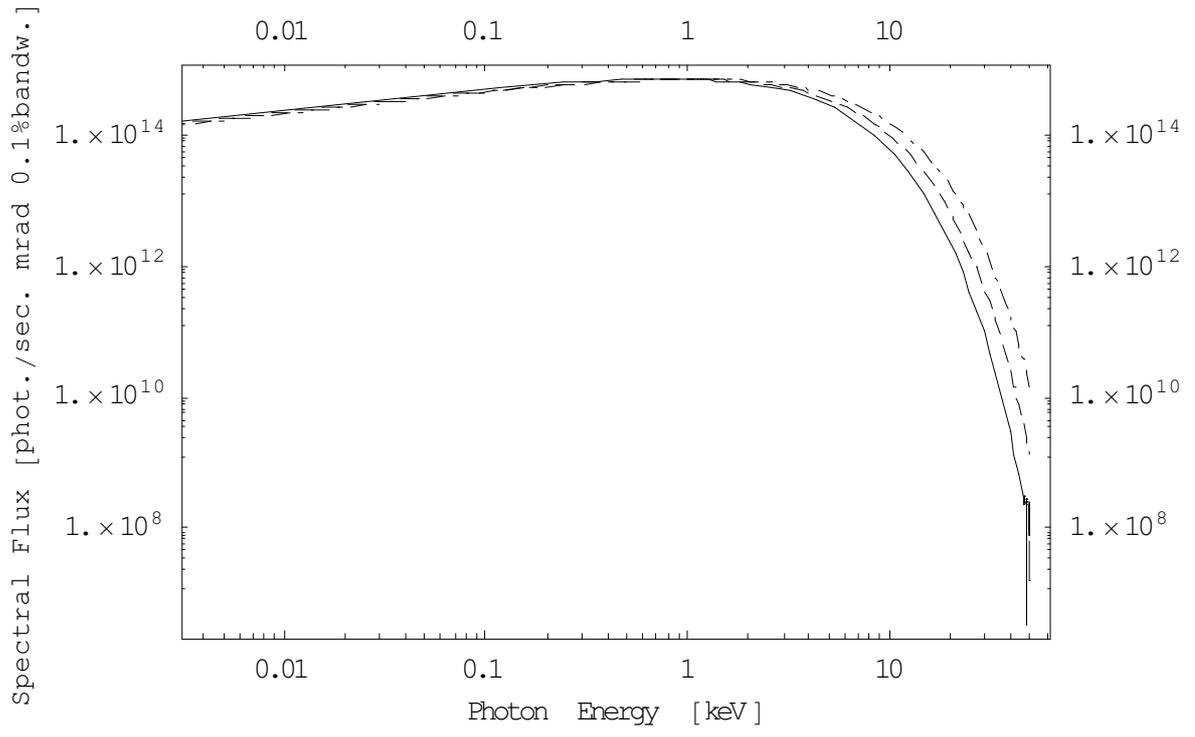
\begin{figure}[tbp] \centering
\caption{Spectral flux of the
SR. Solid line: g=30 mm, Dashed line: g=25 mm, Dot-dashed line: g=20 mm.\label{fig.k.3}}%
\end{figure}

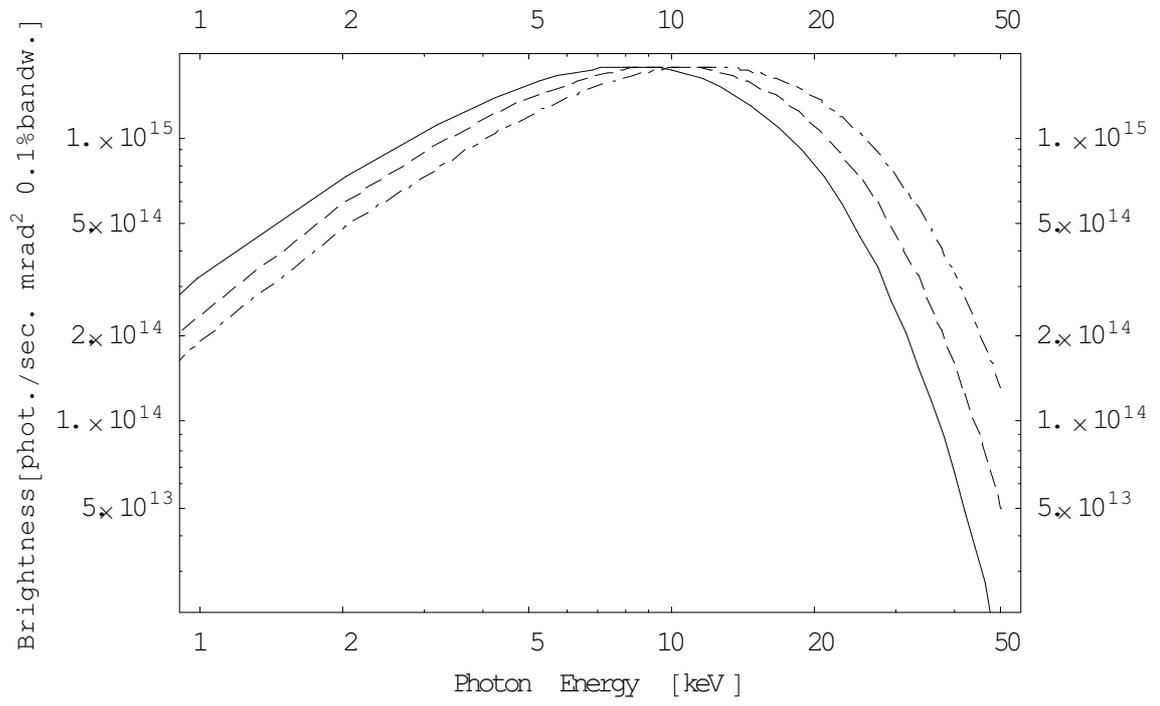
\begin{figure}[tbp] \centering
\caption{Spectral central
brightness of the SR. Solid line: g=30 mm, Dashed line: g=25 mm, Dot-dashed line: g=20 mm.\label{fig.k.4}}%
\end{figure}

Spectral flux and spectral central brightness are given by 
\begin{equation}
I_F[phot./\sec \cdot mrad\cdot 0.1\%bandw]=2.458\cdot 10^{10}\cdot 2N\cdot
I_{e^{+}}[mA]\cdot E_{e^{+}}[GeV]\cdot \frac E{E_c}\int\limits_{E/E_c}^%
\infty K_{5/3}(\eta )d\eta
\end{equation}
and 
\begin{equation}
I_B[phot./\sec \cdot mrad^2\cdot 0.1\%bandw]=1.325\cdot 10^{10}\cdot 2N\cdot
I_{e^{+}}[mA]\cdot E_{e^{+}}^2[GeV]^2\cdot (\frac E{E_c})^2K_{2/3}(\frac E{%
E_c}),
\end{equation}
where $E$ and $E_{c\text{ }}$are photons energy and critical energy,
respectively. Fig. \ref{fig.k.3} and Fig. \ref{fig.k.4} present the spectral
flux and central brightness with respect to photon energy for three
different g values. Critical photon energy is defined by 
\begin{equation}
E_c[keV]=\hbar \omega _c=0.665\cdot (E_{e^{+}}[GeV])^2\cdot B[T]
\end{equation}
Main parameters of SR facility for two options are given in Table \ref
{table.k.1}.

\section{FREE ELECTRON\ LASER FACILITY}

Free Electron Laser (FEL) is a mechanism to convert some part of the kinetic
energy of relativistic electron beam into tunable, highly bright and
monochromatic coherent photon beam by using undulators inserted in linear
accelerators or synchrotrons \cite{pel}. Relativistic electron beam
oscilates on a sinusodial path with the help of a undulator magnet which has
an oscillating magnetic field between its poles. As a result FEL beam is
produced (see, Fig.6.)

Wavelength of the obtained FEL beam is dependent on energy of electron beam,
period of undulator poles and undulator's K parameter

\begin{equation}
\lambda _{FEL}=\frac{\lambda _p}{2\gamma _e^2}(1+\frac{K^2}2)
\end{equation}
where $\lambda _p$ is period length of undulator, $\gamma _e$ is Lorentz
factor of the electron beam. Undulator parameter, K, is given by Eqn. (7).
For undulator $K\simeq 1$ and especially the first harmonic contributes into
the radiation. Laser wavelength and energy for plane undulator in terms of
practical units are given as

\begin{equation}
\lambda _{FEL}[\AA ]=13.056\frac{\lambda _p[cm]}{(E_{e^{-}}[GeV])^2}(1+\frac{%
K^2}2)
\end{equation}
and

\begin{equation}
E_{FEL}[eV]=950\frac{(E_{e^{-}}[GeV])^2}{\lambda _p[cm](1+\frac{K^2}2)}
\end{equation}

Main parameters of FEL\ facility for two options are given in Table \ref
{table.o.1}.

Magnetic field strength between poles of plane undulators is given by Eqn.
(8). For using thereafter, magnetic field is estimated to be 1.48 kG with $%
b=5.47$ , $c=1.8$ , $\lambda _p=33$ $mm$ and $g=25$ $mm.$ With these values,
strength parameter of undulator can be obtained from Eqn.(7) as $K=0.456.$

Flux of FEL\ beam as a function of energy is given as follows \cite{cas}

\begin{equation}
I_{FEL}=1.74\cdot 10^{14}N^2(E_{e^{-}}[GeV])^2\text{ }I\text{ }[A]\text{ }%
F_n[K]\text{ }f(n\nu _n)
\end{equation}

where

\begin{equation}
F_n[K]=\xi n^2[J_{(\frac{n-1}2)}(n\xi )-J_{(\frac{n+1}2)}(n\xi )]^2,\text{ }%
\xi =\frac 12\frac{K^2}{1+K^2}
\end{equation}

and

\begin{equation}
f(\nu )=(\frac{\sin \nu /2}{\nu /2})^2,\nu _n=2\pi N\frac{n\omega _1-\omega 
}{n\omega _1},n=1,3,5...
\end{equation}

Here $J_n$ is n-th order cylindirical Bessel function, $\omega
_1=E_{FEL}/\hbar $ is the frequency of the first harmonic radiation, $N$ is
the number of undulator poles and n is the order of harmonics. Fig. 7 shows
the dependence of FEL\ flux on photon energy for $E_{e^{-}}=250MeV$ option.
Here peaks are placed at odd harmonics and maximum values of fluxes are $%
7.56\cdot 10^{13}$, $1.08\cdot 10^{13}$and $9.45\cdot 10^{11}$ for $n=1,3$
and $5,$ respectively. Obtained averaged brightness values of photon beam
are given in Table IV.

\section{APPLICATIONS\ FIELDS\ OF\ SYNCHROTRON\ RADIATION\ AND\ FREE\
ELECTRON\ LASERS}

Synchrotron radiation sources and free electron lasers have a rich spectrum
of applications (see, for example, \cite{sink}) both in scientific
researches and industry. Part of them are listed below:

Atomic and molecular spectroscopy,

Spectroscopy of atomic and molecular clusters,

Solid state spectroscopy,

Physics and chemistry of surfaces and thin films,

Photochemical processes,

Biological structure and dynamics,

Materials and surface processing,

Multilayer magnetic films,

The electronic structure of semiconductors$,$

Heavy fermion materials and high temperature superconducters,

Dynamics of catalytic reactions.

Proposed SR source will cover a photons wavelengths $\lambda \geq 0.1\AA $,
whereas FEL will produce a laser beam with $\lambda \approx $ ($760-3000)\AA 
$. Due to appropriate modifications of wiggler or undulator parameters these
regions may be extended. Moreover there is also possible option of inserting
undulator in the positron ring to obtaining FEL beam. All these topics will
be considered in forthcoming publications.

\section{CONCLUSION}

In this paper we show that sufficiently high luminosities can be achieved at
linac-ring type $\phi $ factory. Then, the proposed complex will give
opportunity to make a wide spectrum of applied and technological researches.
In this sense linac-ring type $\phi $ factory should be considered as
candidate for the first step of National Acclerator Laboratory due to its
compactness and usefullness.

\section{Acknowledgements}

This work is supported by Turkish State Planning Organization under the
Grant No DPT-97K-120420.

\bigskip\ 

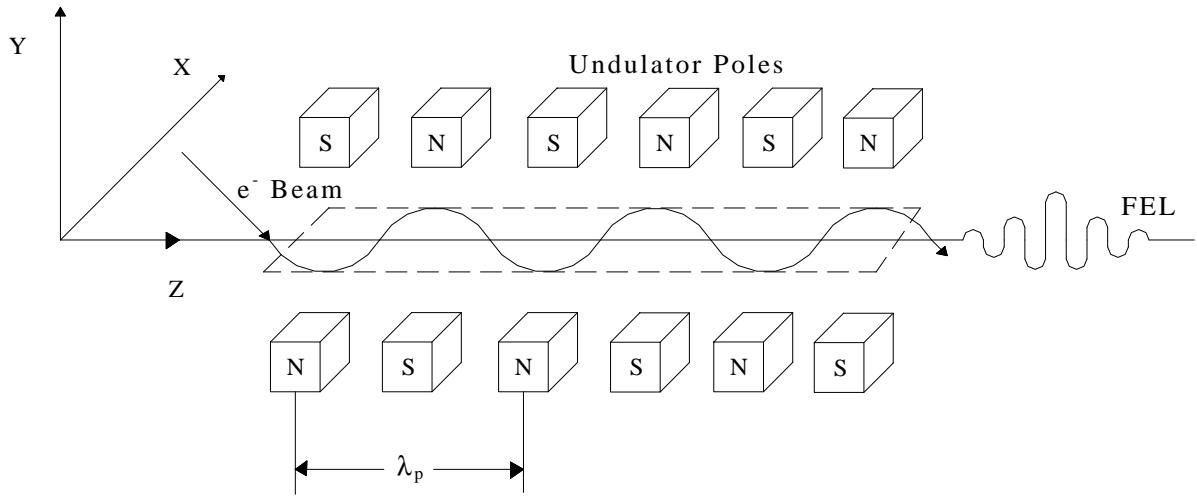
\begin{figure}[tbp] \centering
\caption{Schematic view of FEL
process.\label{fig.o.1}}%
\end{figure}

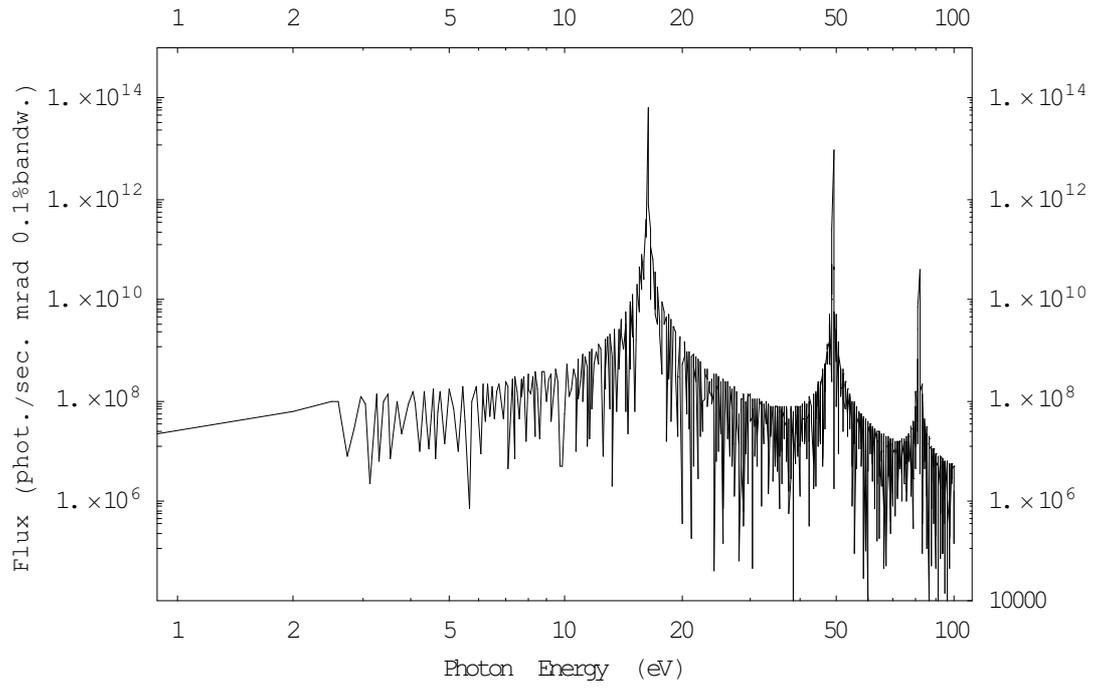
\begin{figure}[tbp] \centering
\caption{Flux of FEL
beam.\label{fig.o.1}}%
\end{figure}

\newpage\ 
\begin{table}[ptb] \centering
\caption{Main parameters of the
$\phi $ factory\label{table.s.1}} 
\begin{tabular}{lll}
Electron beam energy (MeV) & 125 & 250 \\ 
Pozitron beam energy (MeV) & 2000 & 1000 \\ 
Center of mass energy (MeV) & 1000 & 1000 \\ 
Radius of ring (m) & 50 & 30 \\ 
Acceleration gradient (MV/m) & 12.5 & 12.5 \\ 
Lenght of main linac (m) & 10 & 20 \\ 
Number of particles in electron beam ($10^{10}$) & 0.04 & 0.02 \\ 
Number of particles in positron beam ($10^{10}$) & 10 & 20 \\ 
Collision frequency, f (MHz) & 30 & 30 \\ 
Number of bunches in ring, k & 32 & 19 \\ 
Electron current (mA) & 1.92 & 0.96 \\ 
Positron current (A) & 0.96 & 0.48 \\ 
Energy loss /turn, $\bigtriangleup E_{e^{+}}$ (MeV) & 0.03 & 0.003 \\ 
Fractional energy loss of the electrons, $\delta $ (10$^{-4}$) & 2 & 1 \\ 
Beam size at the collision point, $\sigma _{x,y}$ $(\mu m)$ & 1 & 1 \\ 
Beta function at IP, $\beta _{x,y}$ (cm) & 0.25 & 0.25 \\ 
Bunc length, $\sigma _{z\text{ }}$ (cm) & 0.1 & 0.1 \\ 
Luminosity, L ($10^{34}cm^{-2}s^{-1})$ & $\simeq 1$ & $\simeq 1$%
\end{tabular}
\end{table}

\begin{table}[ptb] \centering
\caption{Comparision of event numbers  in DA$\Phi $NE and proposed
collider\label{table.s.2}} 
\begin{tabular}{llll}
Decay chanels & Branching ratios & N (DA$\Phi $NE) & N (LR type $\phi $
Factory) \\ 
$K^{+}K^{-}$ & 0.495 & 1.1$\cdot $10$^{10\text{ }}$ & 2.2$\cdot $10$^{11}$
\\ 
$K_L^0K_s^0$ & 0.344 & 7.6$\cdot $10$^9$ & 1.5$\cdot $10$^{11}$ \\ 
$\rho \pi $ & 0.155 & 3.4$\cdot $10$^9$ & 6.9$\cdot $10$^{10}$ \\ 
$\eta \gamma $ & 1.26$\cdot $10$^{-2}$ & 2.8$\cdot $10$^8$ & 5.6$\cdot $10$%
^9 $ \\ 
$\pi ^0\gamma $ & 1.31$\cdot $10$^{-3}$ & 2.9$\cdot $10$^7$ & 5.8$\cdot $10$%
^8$ \\ 
$e^{+}e^{-}$ & 2.99$\cdot $10$^{-4}$ & 6.6$\cdot $10$^6$ & 1.3$\cdot $10$^8$
\\ 
$\mu ^{+}\mu ^{-}$ & 2.5$\cdot $10$^{-4}$ & 5.6$\cdot $10$^6$ & 1.1$\cdot $10%
$^8$ \\ 
$\eta e^{+}e^{-}$ & 1.3$\cdot $10$^{-4}$ & 2.9$\cdot $10$^6$ & 5.8$\cdot $10$%
^7$ \\ 
$\pi ^{+}\pi ^{-}$ & 8$\cdot $10$^{-5}$ & 1.8$\cdot $10$^6$ & 3.6$\cdot $10$%
^7$ \\ 
$\eta ^{,}$(958)$\gamma $ & 1.2$\cdot $10$^{-4}$ & 2.7$\cdot $10$^6$ & 5.4$%
\cdot $10$^7$ \\ 
$\mu ^{+}\mu ^{-}\gamma $ & 2.3$\cdot $10$^{-5}$ & 5.1$\cdot $10$^5$ & 1.0$%
\cdot $10$^7$%
\end{tabular}
\end{table}

\begin{table}[tbp] \centering
\caption{Main parameters of SR
facility\label{table.k.1}} 
\begin{tabular}{lll}
Energy (GeV) & 1 & 2 \\ 
Maximum magnetic field ($T$) & 1.054 & 1.054 \\ 
Current (A) & 0.976 & 0.488 \\ 
Period (cm) & 13.2 & 13.2 \\ 
Gap (mm) & 30 & 30 \\ 
Total length (m) & 2.112 & 2.112 \\ 
Total radiated power (kW) & 1.44 & 2.90 \\ 
Critical energy, $E_c$(keV) & 0.700 & 2.804 \\ 
Wiggler parameter & 12.99 & 12.99 \\ 
Spectral flux ($Phot/s\cdot mrad\cdot 0.1\%bandw)$ & 4.96$\times 10^{14}$ & 
5.00$\times 10^{14}$ \\ 
Spectral central brightness ($Phot/s\cdot mrad^2\cdot 0.1\%bandw$) & 4.95$%
\times 10^{14}$ & 9.98$\times 10^{14}$%
\end{tabular}
\end{table}

\begin{table}[tbp] \centering
\caption{Main parameters of FEL
facility.\label{table.o.1}} 
\begin{tabular}{|l|l|l|}
\hline
& $E_e=125MeV$ & $E_e=250MeV$ \\ \hline
Photon energy $(eV)$ & 4.07 & 16.30 \\ \hline
Laser wavelength $(\AA )$ & 3044 & 761 \\ \hline
Beam current $(mA)$ & $1.92$ & $0.96$ \\ \hline
Particle per bunch $(10^{10})$ & $0.04$ & $0.02$ \\ \hline
Repetition frequency $(MHz)$ & $30$ & $30$ \\ \hline
Averaged laser beam power $(W)$ for $L=10$ $m$ & $4.18\cdot 10^{-3}$ & $%
8.36\cdot 10^{-3}$ \\ \hline
Flux $(Phot/(s\cdot mrad\cdot 0.1\%bandw))$ & $3.78\cdot 10^{13}$ & $%
7.56\cdot 10^{13}$ \\ \hline
Averaged brightness $(Phot/(s\cdot mrad^2\cdot 0.1\%bandw)$ & $2.91\cdot
10^{11}$ & $5.81\cdot 10^{11}$ \\ \hline
\end{tabular}
\end{table}

\end{document}